\documentclass{article}

\usepackage{arxiv}
\usepackage[utf8]{inputenc}
\usepackage[T1]{fontenc}
\usepackage{hyperref}
\usepackage{url}
\usepackage{booktabs}
\usepackage{amsfonts}
\usepackage{nicefrac}
\usepackage{microtype}
\usepackage{graphicx}
\usepackage{natbib}
\usepackage{doi}

\graphicspath{{pic/}} 

\title{An Integrated UVM-TLM Co-Simulation Framework for RISC-V Functional Verification and Performance Evaluation}

\date{} 

\author{
  Ruizhi Qiu\\
  School of Integrated Circuits Science and Engineering \\
  Beihang University, Beijing, China\\
  \texttt{qiuruizhi02@buaa.edu.cn} \\
  \And
  Yang Liu\\
  School of Integrated Circuits Science and Engineering \\
  Beihang University, Beijing, China\\
  \texttt{liuyang\_me@buaa.edu.cn} \\ 
}

\renewcommand{\shorttitle}{Integrated UVM-TLM Co-Simulation for RISC-V}
\hypersetup{
pdftitle={An Integrated UVM-TLM Co-Simulation Framework for RISC-V Functional Verification and Performance Evaluation},
pdfsubject={Computer Science, Hardware Verification, RISC-V, UVM, TLM},
pdfauthor={Qiu Ruizhi, Liu Yang},
pdfkeywords={RISC-V, Processor Modeling, UVM, TLM, Co-verification, Functional Verification, Performance Evaluation, Co-simulation},
colorlinks=true, linkcolor=blue, citecolor=blue, urlcolor=blue
}

\begin{document}
\maketitle

\begin{abstract}
The burgeoning RISC-V ecosystem necessitates efficient verification methodologies for complex processors. Traditional approaches often struggle to concurrently evaluate functional correctness and performance, or balance simulation speed with modeling accuracy. This paper introduces an integrated co-simulation framework leveraging Universal Verification Methodology (UVM) and Transaction-Level Modeling (TLM) for RISC-V processor validation. We present a configurable UVM-TLM model (\textit{vmodel}) of a superscalar, out-of-order RISC-V core, featuring key microarchitectural modeling techniques such as credit-based pipeline flow control. This environment facilitates unified functional verification via co-simulation against the Spike ISA simulator and enables early-stage performance assessment using benchmarks like CoreMark, orchestrated within UVM. The methodology prioritizes integration, simulation efficiency, and acceptable fidelity for architectural exploration over cycle-level precision. Experimental results validate functional correctness and significant simulation speedup over RTL approaches, accelerating design iterations and enhancing verification coverage.
\end{abstract}


\section{Introduction}
\label{sec:introduction}
The rapid proliferation of the RISC-V Instruction Set Architecture (ISA) \cite{patterson2017computer} offers design flexibility but also presents significant verification challenges, especially for complex, customizable processors incorporating features like superscalar and out-of-order execution \cite{ahmadi2021constrained}. Traditional Register-Transfer Level (RTL) simulations, while cycle-accurate, are prohibitively slow for agile development \cite{kabylkas2021effective, kim2016strober}. High-level architectural simulators such as gem5 \cite{binkert2011gem5} or QEMU \cite{chylek2009collecting} offer speed but may lack microarchitectural detail or seamless integration with industry-standard verification flows like the Universal Verification Methodology (UVM) \cite{przigoda2015verification, diaz2021enabling}. This often leads to a decoupling of functional verification and performance analysis \cite{bhatnagar2019product}.

To address these limitations, this paper proposes an integrated co-simulation framework based on the synergy of UVM \cite{height2012practical, harshitha2021introduction} and Transaction-Level Modeling (TLM) \cite{rath2014transaction, cai2003transaction}. UVM provides a structured, reusable infrastructure for comprehensive functional verification \cite{ni2015research}, while TLM accelerates simulation through abstracted communication \cite{abid2011design}. Our framework utilizes these to create a platform where a detailed RISC-V processor model undergoes simultaneous functional validation against a golden reference (Spike ISA simulator) and performance characterization, all within a cohesive UVM environment. The core of our contribution is this \textbf{integrated verification and evaluation methodology and its enabling framework}, which emphasizes a balance between simulation throughput and acceptable modeling fidelity for early-stage architectural exploration.

The main contributions are:
\begin{itemize}
    \item The design and implementation of an integrated UVM-TLM co-simulation framework tailored for RISC-V processor verification, facilitating concurrent functional validation and performance trend analysis.
    \item A detailed UVM-TLM model (\textit{vmodel}) of a superscalar, out-of-order RISC-V core, demonstrating key modeling techniques for complex microarchitectural features, including a credit-based backpressure flow control for pipeline stages.
    \item Experimental validation of the framework's effectiveness in achieving high functional coverage, identifying microarchitectural issues early, and providing significant simulation speedup compared to RTL approaches, while enabling relative performance comparisons for design space exploration.
\end{itemize}
The remainder of this paper is organized as follows: Section \ref{sec:background_related_work} briefly covers background technologies and related work. Section \ref{sec:framework} details the proposed framework and modeling aspects. Section \ref{sec:methodology} outlines the verification and evaluation procedures. Section \ref{sec:results} presents experimental results. Finally, Section \ref{sec:conclusion} concludes and suggests future work.

\section{Background and Related Work}
\label{sec:background_related_work}
Our approach leverages UVM and TLM. UVM is the industry standard for SystemVerilog-based functional verification, promoting modularity and reusability \cite{height2012practical}. TLM, particularly TLM-2.0 \cite{osci:tlm2req:2007}, abstracts inter-module communication to transactions, significantly speeding up system-level simulations \cite{cai2003transaction}. The synergy of UVM and TLM has been recognized for creating high-performance verification environments, especially for complex SoCs \cite{donlin:codesisss:2004, yun2011beyond}.

Existing RISC-V verification efforts include formal methods \cite{Weingarten2024Complete}, dedicated test generators, and UVM-based functional testbenches \cite{redi_riscv-dv_2019}. Performance modeling often relies on tools like gem5 \cite{Ta2018Simulating} or custom SystemC/TLM models, typically separate from UVM flows. While UVM/TLM integration has been applied to SoC interconnects \cite{samanta2014uvm} or cache coherence \cite{Biswal2017Cache}, a comprehensive UVM-TLM framework for a complex, out-of-order RISC-V core model that deeply integrates functional verification with early performance analysis and detailed microarchitectural modeling techniques (like realistic pipeline flow control) is less explored. Our work aims to address this by providing a unified and efficient solution.

\section{Proposed UVM-TLM Co-Simulation Framework}
\label{sec:framework}
The core of our work is a UVM-TLM based framework for concurrent functional verification and performance assessment of RISC-V processors. It features a detailed, configurable \textit{vmodel} of a superscalar, out-of-order RISC-V core.


\subsection{Framework Architecture}
\begin{figure}[htbp]
  \centering
  \includegraphics[width=0.6\textwidth]{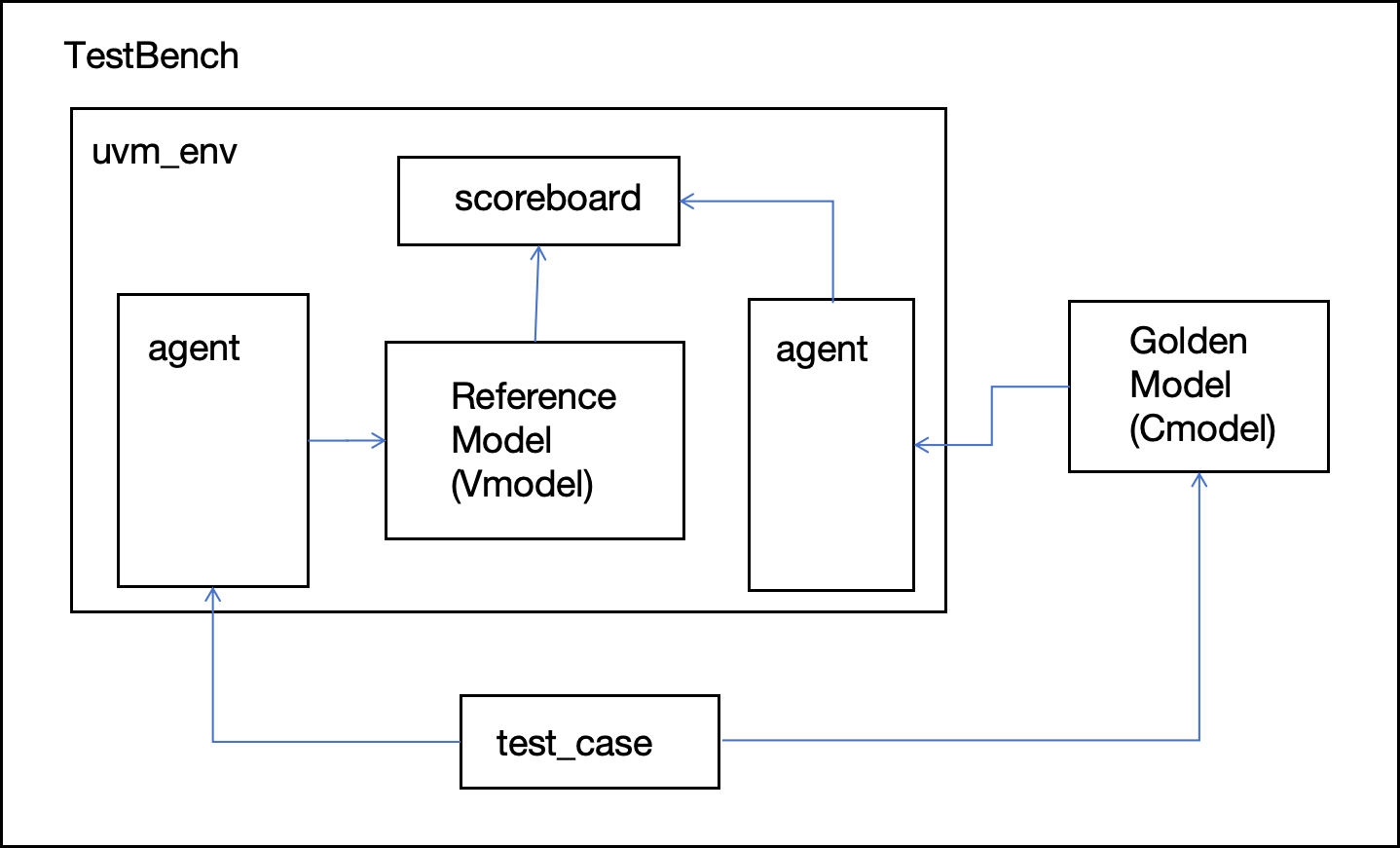}
  \caption{Architecture of the UVM-TLM Co-Simulation Framework.}
  \label{fig:framework_arch}
\end{figure}

The framework (Figure \ref{fig:framework_arch}) uses a layered UVM architecture integrating the \textit{vmodel}, a golden reference model (\textit{cmodel} - Spike \cite{patterson2017computer} via DPI-C), and UVM verification components (testbench, environment, agents, sequencer, comparer/scoreboard).

\subsection{Key Modeling Aspects of \textit{vmodel}}
The \textit{vmodel} is not a black box but reflects key microarchitectural concepts:
\begin{itemize}
    \item \textbf{Modular TLM-Based Communication:} Internal blocks (e.g., Frontend, Backend, Load/Store Unit (LSU), L1 Data cache) communicate via standard UVM-TLM 2.0 interfaces. Instructions, memory requests, and control signals are encapsulated as TLM transactions.
    \item \textbf{Abstracted Pipeline Modeling with Credit-Based Flow Control:} The \textit{vmodel} simulates key pipeline stages (fetch, decode, rename, dispatch, issue, execute, commit). Crucially, to manage transaction flow between pipelined stages (e.g., Frontend to Backend) in the asynchronous TLM environment, a credit-based backpressure protocol is implemented. Downstream modules grant "credits" to upstream modules via dedicated TLM channels, indicating available buffer space. An upstream module consumes a credit upon sending a transaction and stalls if no credits are available. This models realistic pipeline backpressure due to resource contention, enhancing behavioral accuracy beyond simple sequential execution.

    \begin{figure}[htbp]
     \centering
     \includegraphics[width=0.5\textwidth]{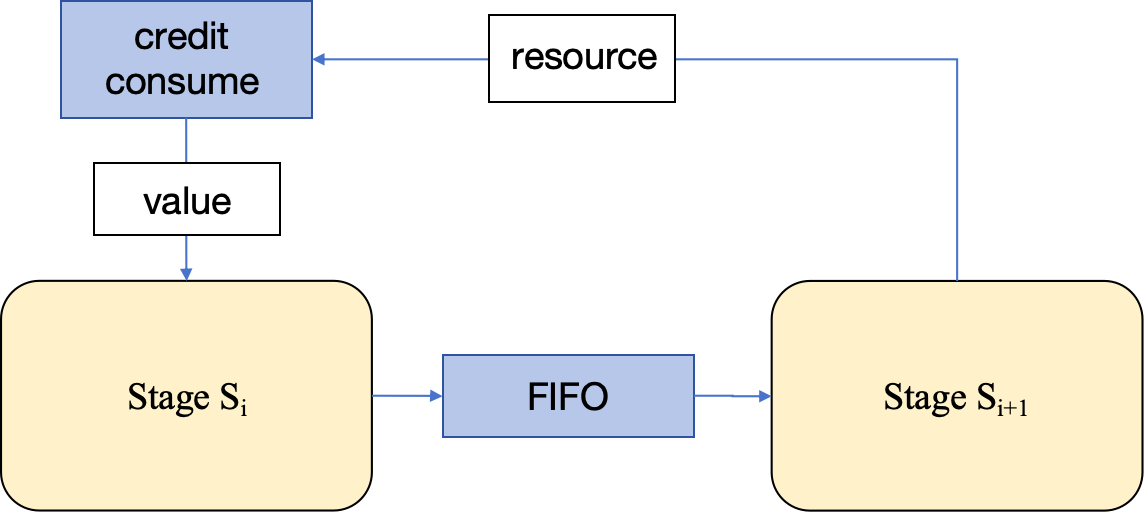} 
     \caption{Conceptual diagram of credit-based flow control between pipeline stages $S_i$ and $S_{i+1}$.}
     \label{fig:credit_flow}
    \end{figure}
    \item \textbf{Integrated Functional and Performance Representation:} TLM transactions can carry approximate timing. The model tracks resource utilization and, importantly, counts retired instructions at the commit stage for accurate performance metrics.
    \item \textbf{Hierarchical and Configurable Structure:} The SystemVerilog codebase is logically organized. The \textit{vmodel} is configurable via UVM's configuration database, allowing parameters like issue width or buffer sizes to be varied for architectural exploration.
\end{itemize}
The frontend (instruction fetch, decode) and backend (rename, dispatch, issue, execution units, commit) are modeled as distinct UVM components interacting via TLM. The LSU handles memory operations, interfacing with an L1D cache model, which in turn connects to a main memory model, all using TLM.

\subsection{Co-Simulation Verification Mechanism}
\begin{figure}[htbp]
  \centering
  \includegraphics[width=0.75\textwidth]{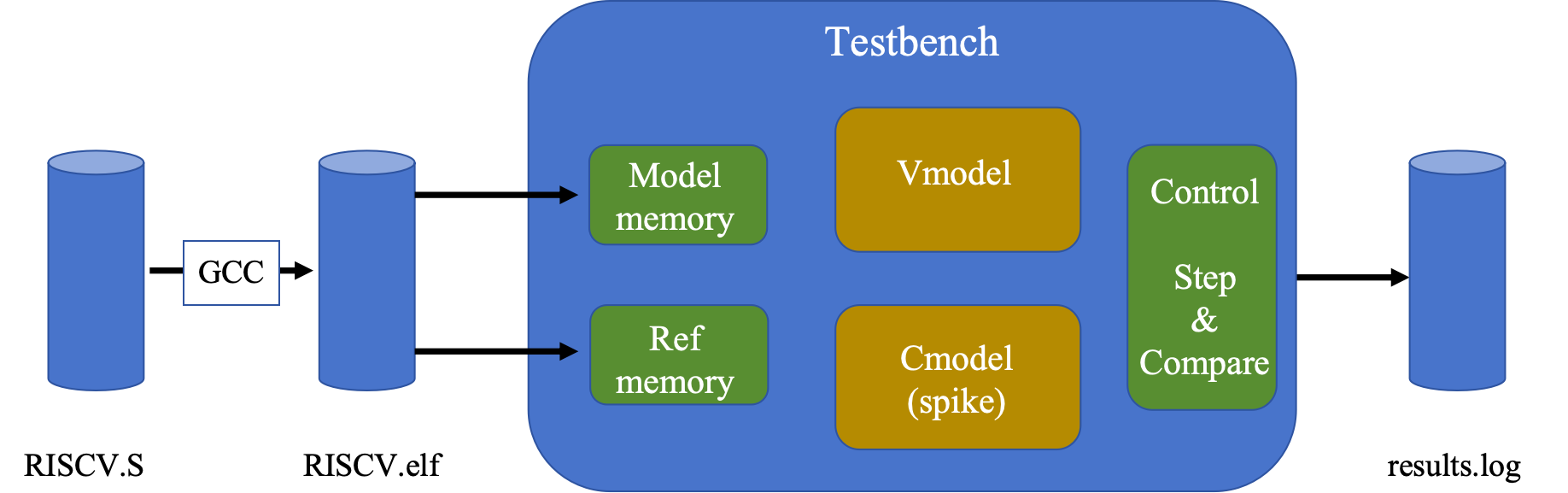}
  \caption{Functional verification flow using co-simulation.}
  \label{fig:cosim_diag}
\end{figure}

Functional correctness is verified by co-simulating \textit{vmodel} and \textit{cmodel} with identical instruction streams (Figure \ref{fig:cosim_diag}). The UVM Comparer module performs a lock-step comparison of architectural states (PC, GPRs, CSRs) upon instruction commitment in \textit{vmodel}, flagging any discrepancies.

\section{Integrated Verification and Evaluation Methodology}
\label{sec:methodology}
The framework enables a cohesive workflow for both functional verification and performance evaluation.
\subsection{Functional Verification Procedure}
Standardized test suites (e.g., \textit{riscv-tests}) are compiled for the \textit{vmodel}'s target ISA. Binaries are loaded into a shared memory model (potentially via backdoor access for speed). The UVM testbench drives execution on both \textit{vmodel} and \textit{cmodel}, with the Comparer automatically performing state comparison upon instruction retirement.

\subsection{Performance Evaluation Procedure}
Using the same UVM environment, benchmarks like CoreMark \cite{gal2012exploring} are run on \textit{vmodel}. Key Performance Indicators (KPIs) like Instructions Per Cycle (IPC) are derived from:
\begin{itemize}
    \item \textbf{Execution Cycles:} Precisely tracked by the UVM environment for the \textit{vmodel} core.
    \item \textbf{Retired Instructions:} Accurately counted within \textit{vmodel} as instructions are architecturally committed in the final pipeline stage, providing a reliable measure of useful work. This count is propagated to the UVM environment.
\end{itemize}
A dedicated UVM test sequence manages the performance run, including configuring microarchitectural parameters of \textit{vmodel}, loading the benchmark, initiating/halting counters, and calculating KPIs (e.g., IPC = Retired Instructions / Execution Cycles).

\subsection{Methodological Emphasis and Fidelity}
The primary objective is \textbf{integrated assessment and relative performance analysis}. The UVM-TLM model, by design (e.g., approximate timing, simplified L1 cache model), does not yield cycle-accurate results comparable to RTL. Its value lies in significant simulation speedup, enabling rapid evaluation of \textit{performance trends and the relative impact} of different microarchitectural configurations (e.g., varying issue width, execution unit count) early in the design phase.

\section{Results and Discussion}
\label{sec:results}
Experiments were conducted using Synopsys VCS on a Linux server environment with Intel Xeon Gold CPUs.

\subsection{Functional Verification Efficacy}
The \textit{vmodel} (configured for RV64IMAFDC) successfully passed all relevant \textit{riscv-tests} (Table \ref{tab:riscv_tests_results}), confirming ISA-level functional correctness.
\begin{table}[htb]
  \centering
  \caption{RISC-V Compliance Test Suite (\textit{riscv-tests}) Execution Summary}
  \label{tab:riscv_tests_results}
  \begin{tabular}{lccc}
    \toprule
    \textbf{Test Subset} & \textbf{Target ISA} & \textbf{Executed Cases} & \textbf{Pass Rate} \\
    \midrule
    RV64I            & RV64IMAFDC &   81   &  100\%  \\
    M, A, F, D, C Ext. & RV64IMAFDC & 97 & 100\% \\ 
    \bottomrule
  \end{tabular}
\end{table}

More significantly, the UVM environment's white-box observability and functional coverage collection (achieving >97\% for key microarchitectural mechanisms) proved crucial for in-depth validation. This approach enabled the detection and diagnosis of subtle microarchitectural bugs that are often masked in purely ISA-level "black-box" comparisons. For instance, through detailed UVM monitoring and coverage analysis of internal pipeline states and data paths, we identified issues such as: (1) an intermittent data error due to data forwarding path contention in specific dual-issue scenarios under high load, and (2) an operand partitioning error during the decoding of certain non-standard instruction variants. Discovering such bugs at the TLM abstraction level, facilitated by UVM's introspection capabilities, offers a significant time saving (estimated 4-6 weeks in our project timeline) compared to their potential discovery much later in the RTL verification phase, thereby greatly improving overall design and verification efficiency.

\subsection{Performance Assessment and Simulation Efficiency}
Performance evaluation was conducted using CoreMark 1.0 (compiled with GCC 9.4.0, -O2). The \textit{vmodel} baseline configuration was a single core, 4-wide issue, 128-entry ROB, 32KB 8-way L1D cache (abstracted), and a BTB+TAGE-like branch predictor. Results are compared to a Xiangshan Nanhu processor reference (Table \ref{tab:coremark_combined_results}).

\begin{table}[htb]
  \centering
  \caption{CoreMark 1.0 Performance Comparison}
  \label{tab:coremark_combined_results}
  \begin{tabular}{lcccc}
    \toprule
    \textbf{Model/System} & \textbf{Retired Instr.} & \textbf{Exec. Cycles} & \textbf{IPC} & \textbf{CoreMark/MHz} \\
    \midrule
    \textit{vmodel} (This Work) & $\approx$24.5 B & $\approx$10.2 B & $\approx$2.40 & $\approx$8.90 \\
    Xiangshan Nanhu (Ref.) & $\approx$31.2 B & $\approx$13.0 B & $\approx$2.40 & $\approx$7.81 \\
    \bottomrule
  \end{tabular}
\end{table}

The \textit{vmodel}'s CoreMark/MHz score exhibits an approximate 14.1\% deviation from the Xiangshan reference. This level of accuracy is within the acceptable range for early-stage TLM models where the primary goal is architectural exploration and relative performance trend analysis, rather than absolute cycle-level precision \cite{osci:tlm2req:2007, pasricha2010chip, kogel:mpsoc:2008}. The deviation primarily stems from abstractions in the TLM model, notably the simplified L1 cache implementation (currently pass-through without detailed hit/miss latency modeling), generalized pipeline stage timing, and idealized resource contention models. Despite these abstractions, the model effectively captures relative performance variations when microarchitectural parameters (e.g., execution unit counts, queue depths) are modified, validating its utility for design space exploration. Some specialized high-accuracy TLM simulations have reported lower error margins \cite{kusoclab:acctlmsim:talks}, indicating avenues for future refinement.

The most significant advantage of the UVM-TLM framework is its \textbf{simulation efficiency}. Executing the full CoreMark benchmark on the \textit{vmodel} took approximately 10-15 minutes on our simulation platform. In stark contrast, an equivalent simulation run on a corresponding RTL model is estimated to require 10-12 hours. This orders-of-magnitude speedup dramatically accelerates design-verify-debug cycles, enabling more extensive testing and rapid exploration of architectural alternatives early in the design process.

\section{Conclusion and Future Work}
\label{sec:conclusion}
This paper presented an integrated UVM-TLM co-simulation framework designed for the efficient verification and early performance evaluation of complex RISC-V processors. Our approach, featuring a detailed UVM-TLM model (\textit{vmodel}) of a superscalar, out-of-order RISC-V core, facilitates concurrent functional validation against the Spike ISA simulator and performance trend analysis using standard benchmarks. Key modeling aspects, such as credit-based pipeline flow control, were discussed.

Experimental results demonstrated the framework's ability to ensure ISA-level functional correctness and, more importantly, leverage UVM's capabilities for in-depth microarchitectural validation, leading to early detection of subtle design issues. While the TLM abstraction level results in a deviation in absolute performance metrics (approx. 14.1\% for CoreMark/MHz against a reference), the model provides valuable relative performance insights and achieves orders-of-magnitude simulation speedup compared to traditional RTL methods. This significantly accelerates architectural exploration and design iteration.

Future work will focus on: (1) \textbf{Enhancing Model Fidelity} by implementing a more detailed L1/L2 cache hierarchy with accurate hit/miss logic and realistic timing, and refining timing models for other critical pipeline components. (2) \textbf{Exploring Model-Driven RTL Development} by investigating how the structured TLM model can guide or partially automate the generation of RTL code. (3) \textbf{Expanding Verification Scope} to include advanced microarchitectural features, multi-core systems, and security aspects.

The proposed UVM-TLM framework offers a practical and effective methodology for navigating the complexities of modern RISC-V processor design and verification.


\bibliographystyle{unsrt} 
\bibliography{ref}  

\end{document}